\documentclass[twocolumn,showpacs,preprintnumbers,amsmath,amssymb,aps,prl]{revtex4}
\usepackage{graphicx}
\begin{document}

\title{Static and Dynamic Phases for Vortex Matter with Attractive Interactions}   
\author{J.A. Drocco, C.J. Olson Reichhardt,
C. Reichhardt, and A.R. Bishop} 
\affiliation{
Center for Nonlinear Studies and
Theoretical Division,
Los Alamos National Laboratory, Los Alamos, New Mexico 87545 USA} 

\date{\today}
\begin{abstract}
Exotic vortex states with long range attraction and short range 
repulsion have recently been proposed to arise in superconducting hybrid structures and 
multi-band superconductors.
Using large scale simulations we examine the static and dynamic properties of such vortex states
interacting with random and periodic pinning.
In the absence of pinning this system does not 
form patterns
but instead 
completely phase separates.
When pinning is present there is a transition from inhomogeneous to homogeneous
vortex configurations similar to a wetting phenomenon.
Under an applied drive, 
a dynamical dewetting process can occur from 
a strongly pinned homogeneous state into pattern forming states. 
We show that a signature of the exotic vortex interactions 
under transport measurements is a robust double peak feature
in the differential conductivity curves.  
\end{abstract}
\pacs{74.25.Wx,74.25.Uv,89.75.Kd}
\maketitle

\vskip2pc
Vortices in superconductors interacting with random or periodic pinning provide
a model system 
for studying the interplay    
between 
particle interactions that favor one type of ordering
and substrate interactions
that favor a disordered state 
or some alternative symmetry \cite{1}.
In 
such systems a rich variety of liquid \cite{1}, 
glassy \cite{27}, 
and incommensurate
static phases \cite{21,22} arise, 
and under an external drive 
numerous dynamical phases \cite{1,23,25,26} 
occur that 
produce distinct 
transport 
measurement 
features 
\cite{28,29,30,23}.
Most vortex systems have strictly repulsive
vortex-vortex interactions;
however, modified interactions (MI) consisting of short range
repulsion and long range attraction 
are proposed to occur in
low-$\kappa$ superconductors \cite{16,5} or 
type-I/type-II hybrid superconducting structures \cite{14}.
Recent imaging experiments in the multi-band superconductor 
MgB$_{2}$ revealed inhomogeneous, stripe-like vortex
configurations 
\cite{7,8},
interpreted as evidence that 
the
multi-band nature of the system produces coexisting type-I and type-II vortices,
giving long range attractive and short range repulsive
vortex-vortex interactions
\cite{3,2,4,7,5,6,9,9a,10,11,12,13,15}.  
There has been some controversy regarding
this interpretation due to the presence of quenched disorder or pinning, since
even for vortices with purely repulsive interactions, strong pinning can 
produce highly inhomogeneous flux distributions \cite{31}. 

Previous work has shown that a variety of periodic clump crystals and stripe
structures can occur in systems with competing
long range repulsive and short range attractive interactions
or with two step repulsive interactions 
\cite{18,20,19,17}.   
It is not clear that such inhomogeneous structures would occur 
for interactions with long range attraction
due to the
low energy of a completely phase separated state.
Numerical simulations of the proposed MI vortices
found evidence 
of clump or labyrinth structures \cite{24,24a}; 
however, other studies
of this system found complete phase separation \cite{15}. 
In order to address the ground
state ordering of the 
MI vortex
system 
with and without quenched disorder, 
we perform large scale numerical simulations. 
We find that for long simulated annealing 
times 
and no pinning,
pattern formation 
does not occur and 
the vortices completely phase separate
into a single giant clump, while
for shorter annealing times
the system can become 
trapped in metastable
states exhibiting some pattern formation. 
The addition of pinning induces a crossover from the phase
separated state to a homogeneous,
pattern-free vortex state. 
This behavior is very similar to the wetting-dewetting transition 
found for 
liquids of mutually attractive atoms on surfaces.
When the atomic attraction dominates, the atoms clump or form a single
dewetted drop, while when the surface attraction dominates, the atoms
spread out and wet the surface \cite{32}. 
We find that
under an applied drive, 
for strong pinning
the driven states can form patterns
such as stripes or labyrinths.
We show that transport measurements provide
a clear way to 
determine whether 
MI vortices exist in a sample,
since the dynamic phase transition into inhomogeneous 
clump or stripe states 
produces
a robust double peak in the differential conductivity. 
Our results are general
and can be applied to other systems of particles 
with competing long range attraction and
short range repulsion in the presence of a 
substrate,
such as colloidal particles
or vortices in Bose-Einstein condensates \cite{33}.  

{\it Simulation--}
We simulate a two-dimensional (2D) system of 
$N_{v}=400$ vortices and $N_{p}$ pinning sites 
with periodic boundary conditions in the $x$ and $y$ directions. 
The vortex dynamics are obtained by integrating the following equation of 
motion:
$\eta (d {\bf R}_{i}/dt) = {\bf F}^{vv}_i + {\bf F}^{vp}_i + {\bf F}_{d} + {\bf F}^{T}_i$,
where $\eta$ is the damping constant and ${\bf R}_{i}$ 
is the position of vortex $i$. 
We use the vortex-vortex interaction force 
proposed for the type-I/type-II hybrid materials and 
multi-band superconductors \cite{6,24}, 
\begin{equation}    
{\bf F}^{vv}_{i} = \sum_{ j \neq i}^{N_v}[aK_{1}(bR_{ij}/\lambda) - K_{1}(R_{ij}/\lambda)]{\bf \hat R}_{ij}.
\end{equation}  
Here $K_{1}(r)$ is the modified Bessel function, 
$R_{ij}=|{\bf R}_i-{\bf R}_j|$, 
${\bf \hat R}_{ij}=({\bf R}_i-{\bf R}_j)/R_{ij}$, and
$\lambda$ and $\lambda/b$ are
the penetration depths in the two bands.
In previous pin-free simulations of this model, 
the constants $a$ and $b$ were varied
to change the relative strength of the attractive term
\cite{24}. 
Long range attraction and short range repulsion are obtained
when $a > b$ and $b > 1.0$.
The coefficient $a=K_{1}(r_c)/K_{1}(br_c)$, and we set
$r_{c} = 2.1\lambda$ and $b = 1.1$. 
Since the interaction falls off rapidly at large distances we cut off
the potential for
$R_{ij} > 8\lambda$. 
We model the pinning sites as attractive parabolic 
traps with maximum force $F_{p}$ and
radius $R_{p} = 0.3\lambda$, as previously used in simulations of
repulsive vortices 
\cite{26,22}.
The vortex to pinning site ratio 
is measured in terms of
$B_{\phi}/B$, where $B_{\phi}$ is the field at 
which there is one vortex per pinning site and $B$ is the vortex density, 
held fixed throughout this work.
The thermal force term ${\bf F}^{T}_{i}$ 
has the following properties: $\langle F_{i}^{T}(t)\rangle = 0.0$
and $ \langle F^{T}(t)F_{i}^{T}(t^{\prime})\rangle = 2\eta k_{B}T\delta_{ij}\delta(t - t^{\prime})$.  
After performing simulated annealing we
apply an external drive ${\bf F}_{d}=F_d{\bf \hat x}$ and
measure the resulting vortex 
velocity $V_x = N^{-1}_{v}\sum^{N_{v}}_{i}{\bf v}_i \cdot {\bf {\hat x}}$, 
which would correspond
to an experimentally measured current-voltage curve.  

We first study the model 
without pinning
and conduct simulated annealing studies by starting
from a high temperature molten state and slowly cooling to $T=0$
during a time $\tau_{a}$.
For instantaneous annealing (small $\tau_{a}$) the vortices 
fall into a disordered
assembly of clumps \cite{24}, 
labyrinths, or voids. 
The structures coarsen as 
$\tau_{a}$ is increased, and for long
$\tau_{a}$,
the system completely phase separates 
into a single clump. 
Such complete phase separation was observed in
Landau-Ginsburg simulations of this same MI vortex model \cite{15}. 
These results suggest that the stripes and inhomogeneous patterns 
imaged in MgB$_{2}$ \cite{7,8}, a weak pinning material, 
are not produced by 
MI vortices.
It is also possible that 
MI vortices are present in this
material but that pinning arrests
the phase separation. 
Even if pinning could
stabilize pattern forming structures,
it is also possible for inhomogeneous pinning to produce similar
structures for vortices with only repulsive interactions, so the 
structures imaged in \cite{7,8}
do not provide conclusive evidence of the nature of the vortex
interaction potential.
               
\begin{figure}
\includegraphics[width=3.5in]{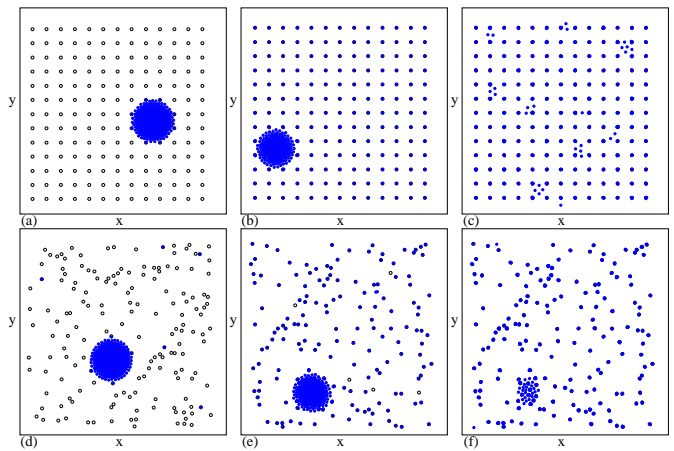}
\caption{
The vortex positions (filled circles) and pinning sites (open circles) 
after annealing
in the presence of periodic (a,b,c) or random (d,e,f) 
pinning for $B_{\phi}/B = 0.4225$. 
(a) Dewetted (D) state 
at $F_{p} = 0.3$
where most pinning sites are unoccupied. 
(b) Partially wetted (PW) state at
$F_{p} = 0.9$. 
Each 
pin captures at least one vortex
and the remaining vortices form a single clump. 
(c) Wetted (W) state at $F_p=1.5$ where most vortices
are trapped by 
pins.
(d) D state 
at $F_{p} = 0.3$. (e) PW state at $F_{p} = 0.9$. 
(f) W state at $F_{p} = 1.5$.}
\label{fig:1}
\end{figure}

\begin{figure}
\includegraphics[width=2.7in]{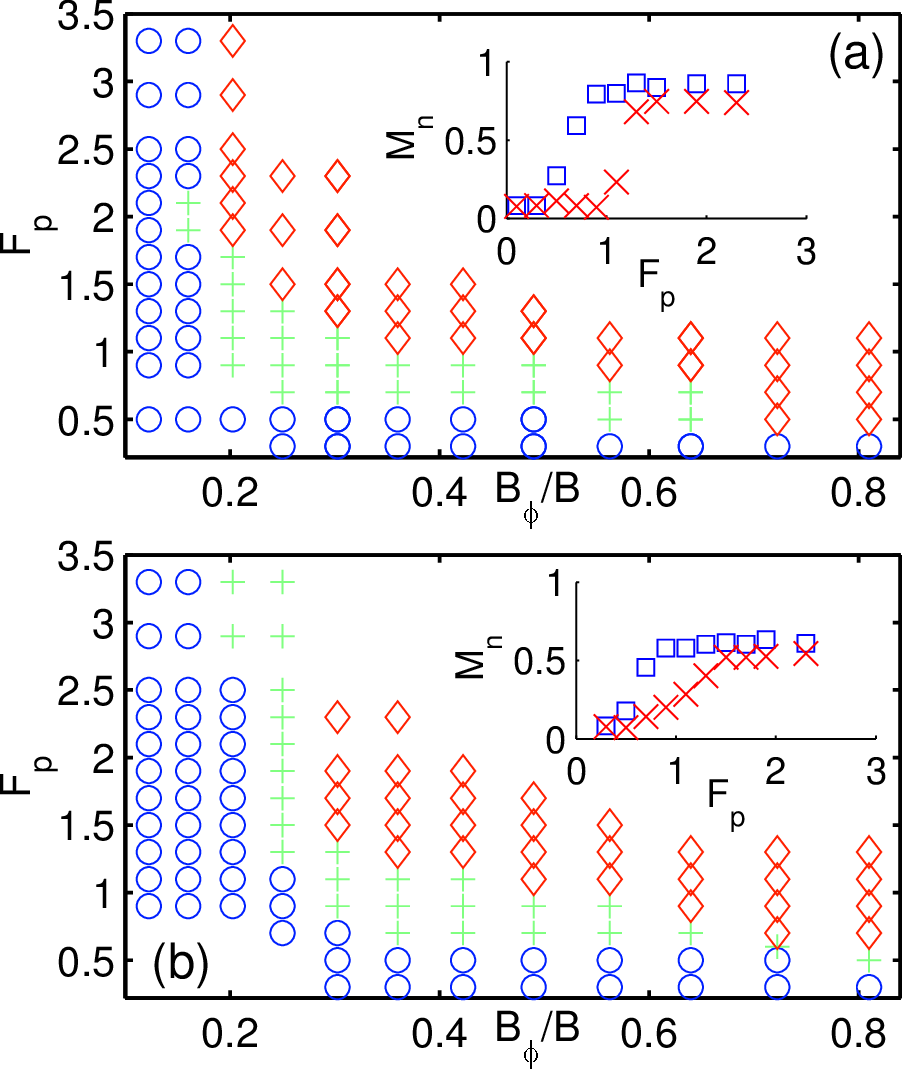}
\caption{ 
$F_p$ vs $B_\phi/B$ phase diagram showing the D ($\bigcirc$), PW (+), and
W ($\diamond$) states for samples with fixed $B$ and varied $B_\phi$.
(a) Periodic pinning. (b) Random pinning.
Insets: $M_n$, the fraction of pins containing at least $n$ vortices,
vs $F_{p}$ for periodic
(upper inset) and random (lower inset) pinning. 
Squares: $M_1$;
x's: $M_2$.
}
\label{fig:diverge}
\end{figure}

In Fig.~1 we show the vortex and pin positions 
from long $\tau_{a}$ 
simulations of samples with periodic and random pinning at
$B_\phi/B=0.4225$.
At $F_p=0.3$ for square pinning,
Fig.~1(a) shows complete phase separation of the vortices
into a single clump, 
with all pins empty except those directly under the clump.
In Fig.~1(b) at $F_{p} = 0.9$,
each 
pin captures one vortex 
producing a homogeneous flux background, while
the remaining vortices form a single clump.
At $F_p=1.5$ in Fig.~1(c), 
each
pin captures multiple vortices, 
and there are patches of unpinned interstitial vortices. 
For larger values of $F_{p}$, all the vortices are trapped
by pins.
Similar behavior occurs for random pinning, as shown in
Fig.~1(d-f).  The single clump state for weak pinning [Fig.~1(d)] is
followed at higher pinning strength by a
phase with a coexisting clump and uniform flux background, shown
in Fig.~1(e) for $F_p=0.9$. 
The clump decreases in size  
as $F_{p}$ increases, and eventually all the vortices become pinned as
shown in Fig.~1(f) for $F_p=1.5$.  We have investigated a range of values
of $F_p$ and $B_\phi/B$ as well as other system parameters, and in general
always find the same generic phases.  We find no regimes where multiple
clump, stripe, or labyrinth phases are stabilized, nor are there
any phases that resemble the structures observed for systems with
long range repulsive and short range attractive interactions
\cite{18,19,20}.
In analogy to the spreading of liquid drops on surfaces \cite{32}, we
term the states illustrated in Fig.~1 dewetted (D), where all the
vortices form a single clump; partially wetted (PW), where all the pins are occupied but there is still a vortex clump; and wetted (W), where the clump
is absent and the majority of the vortices are trapped by pins.

Figure~2 shows the locations of the D, PW, and W regimes in the $F_p$ versus
$B_\phi/B$ phase diagram obtained for 
both periodic and random samples where we 
vary $B_\phi$ by changing the pinning
density.  
To identify 
the locations of the phases we measure
$M_n$, the fraction of pins containing at least $n$ vortices,
as shown in the insets of Fig.~2.  Both $M_1$ and $M_2$ are low in the 
D phase; $M_1$ is high and $M_2$ is low in the PW phase; and $M_1$ and $M_2$
are both high in the W state.
We find that the system enters the W state 
at high $F_p$ and $B_\phi/B$, while for weak $F_p$ or low
$B_\phi/B$, the D state dominates.

\begin{figure}
\includegraphics[width=3.5in]{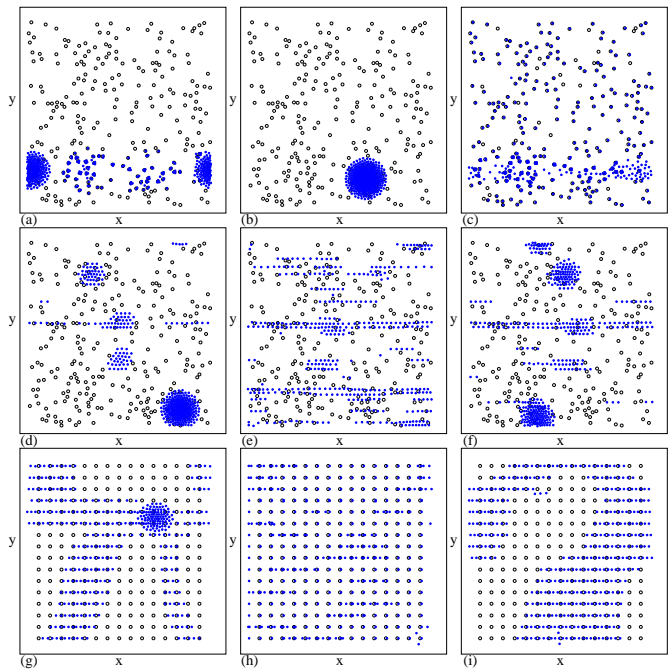}
\caption
{
The vortex positions (filled circles) and pinning sites (open circles)
in snapshots of the moving state for a sample with (a-f) random pinning 
and (g-i) periodic pinning
at $B_\phi/B=0.64$ under a
drive $F_d{\bf \hat x}$
showing the formation of heterogeneous
states at high drives.
(a) $F_{p} = 0.3$, $F_{d} = 0.05$ ($i$).
(b) $F_p=0.3$, $F_d=0.8$ ($ii$).
(c) $F_p=0.7$, $F_d=0.25$ ($iii$).
(d) $F_p=0.7$, $F_d=3.0$ ($iv$).
(e) $F_p=1.3$, $F_d=2.5$ ($v$).
(f) $F_p=1.3$, $F_d=3.8$ ($vi$).
(g) $F_{p} = 0.7$, $F_d=1.6$ ($vii$). 
(h) $F_p=1.3$, $F_d=0.96$ ($viii$). 
(i) $F_p=1.3$, $F_d=3.0$ ($ix$).
The labels $i$-$vi$ [$vii$-$ix$]
correspond to the drives marked in Fig.~4(a-c) [(e,f)].
}
\label{fig:vt}
\end{figure}

\begin{figure}
\includegraphics[width=3.5in]{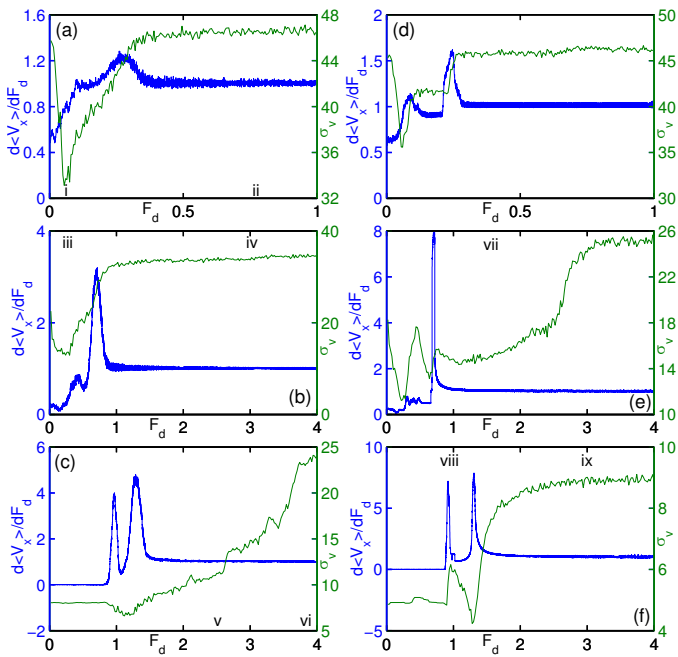}
\caption{ 
$d\langle V_x\rangle/dF_{d}$ (left axes, dark blue curves) and 
$\sigma_v$ (right axes, light green curves) vs $F_{d}$. 
Higher values of $\sigma_v$ correspond 
to more heterogeneous vortex configurations.
(a-c) A sample with random pinning at
$B_{\phi}/B = 0.64$ for (a) $F_{p} =0.3$, (b) $F_{p} = 0.7$,
and (c) $F_{p} = 1.3$. 
The letters $i$-$vi$ indicate the points at which the images in Fig.~3(a-f)
were taken.
(d-f) A sample with periodic pinning at $B_{\phi}/B  = 0.64$ 
for (d) $F_p=0.3$, (e) $F_p=0.7$, and (f) $F_p=1.3$.
The letters $vii$-$ix$ indicate the points at which the images in 
Fig.~3(g-i)
were taken.
}
\label{fig:sixpanel}
\end{figure}

We now show that the most promising method for determining 
whether 
MI vortices are present
in a sample is by using an  
external drive. 
Studies of vortices with strictly repulsive interactions driven over random 
pinning 
find that a disordered pinned state can undergo a dynamical ordering
transition in the moving state when the effective pinning is reduced
\cite{25,26,28,29,30}.   
The onset of  the dynamic ordering 
is associated with
a peak in
the $dV/dI$ curves in experiments \cite{29,28} 
or $d\langle V_x\rangle /dF_d$ curves in simulations \cite{29,26,30}. 
For systems with 
MI vortices,
strongly pinned samples form the uniformly pinned 
state illustrated in Fig.~1(c,f).
A dynamical structural transition occurs 
under an applied drive 
due to the reduced effectiveness of the pinning in the moving state, 
which allows the attractive part of the vortex interactions 
to draw the vortices into a highly heterogeneous configuration.
In Fig.~3 we show snapshots of
the vortex positions in the driven state for random pinning
samples with three different pinning strengths at
$B_{\phi}/B = 0.64$.
In the D state, two types of dynamics occur.
When $F_{p}$ is very weak, the clump 
depins and remains intact; however, for
higher $F_{p}$, above depinning
the clump sheds a trail of pinned vortices,
as shown in Fig.~3(a) 
for $F_{p} = 0.3$ and $F_{d} = 0.05$.
The trailing edge of the clump becomes rarefied,
leading to a decrease in the effective attraction between vortices
that allows them to be torn away from the clump
by the pins.
As $F_{d}$ increases, the trapped vortices depin
and rejoin the clump, which retains its shape for 
higher drives.
The reassembled clump for the $F_p=0.3$ sample
is shown at $F_d=0.8$ in Fig.~3(b).
In Fig.~4(a) we plot $d\langle V_x\rangle/dF_{d}$ and $\sigma_{v}$ versus $F_{d}$ 
for the random pinning sample in Fig.~3(a).  Here
$\sigma_{v}$ is the standard deviation of the vortex density calculated
using a 2D grid with elements equal in length to the average
pinning lattice constant $a$.
Homogeneous vortex configurations give small values of $\sigma_v$, while
$\sigma_v$ is larger
for heterogeneous configurations.
The pinned 
clump state has a large value of $\sigma_v$.
As $F_d$ is increased, $\sigma_v$ initially
drops when the moving clump sheds vortices, as in Fig.~3(a),
and then increases again when the clump reforms for higher drives,
as in Fig.~3(b).
There is a weak two peak structure in 
$d\langle V_x\rangle/dF_d$ reflecting the two step depinning transition,
with 
the second peak corresponding to the drive at which all of the vortices
depin.
Figure~3(c,d) shows a sample with $F_{p} = 0.7$, 
which begins in a PW state.
Above depinning, shown in Fig.~3(c) for $F_{d} = 0.25$, 
the clump disintegrates and the vortices flow
in a meandering path, while at higher drives the vortices
reorder into a much more heterogeneous state containing a clump
and patches of filaments, as shown in Fig.~3(d) at 
$F_{d} = 3.0$.
Figure 4(b) shows that $\sigma_v$ is high in the pinned heterogeneous
state, passes through
a local minimum near the
first peak in $d\langle V_x\rangle/dF_{d}$ when the clump breaks apart,
and then increases again to a value higher than that at $F_d=0$
in the vicinity of the 
second peak in $d\langle V_x\rangle/dF_{d}$ where
the vortices form a moving clump and filament state.
In Fig.~3(e,f) we illustrate the moving states 
at $F_d=2.5$ and $F_d=3.8$ for 
a 
sample with $F_{p} = 1.3$ that has a W pinned state.
The vortices form a moving stripe state near $F_d=2.75$,
and a portion of the stripes
collapse back into clumps as $F_d$ increases; however, even for high $F_d$,
some stripes remain present. 
Figure~4(c) shows that the $d\langle V_x\rangle/dF_d$ curve 
has
a pronounced double peak and that $\sigma_{v}$  
increases with increasing $F_{d}$ as the vortex configuration 
becomes more homogeneous. 
In general, for higher $F_p$ we find more stripe-like driven states, and
the double peak feature in $d\langle V_x\rangle/dF_d$ becomes even
more prominent.
These results show that transport measures are the most useful 
approach for determining
whether MI vortices are present in
multiband or hybrid superconductors, since such 
vortices produce
robust double peak features in the I-V curves.  Complementary imaging
experiments should show 
increasing
heterogeneity of the vortex structure at higher drives. 
We note that the dynamical driven states found here are
much more inhomogeneous than the states found for systems 
with long range repulsive and short range attractive interactions \cite{20}. 

The dynamics for samples with periodic pinning show the same overall trends.
In the D regime, the moving clump 
sheds a trail of vortices above depinning,
and as $F_d$ increases, 
all the vortices 
depin
and the clump reforms.
The corresponding $d\langle V_x\rangle/dF_{d}$ 
in Fig.~4(d) has a double peak feature correlated with
jumps in $\sigma_v$ to higher values 
as the system becomes more heterogeneous at higher $F_d$.
For the PW state,  
above depinning the clumps break apart and
shed both pinned and interstitial vortices.
As $F_d$ increases,
the clumps begin to reform until 
all the vortices depin and form a coexisting clump and moving labyrinth
state with vortices flowing along the pinning rows, as illustrated
in Fig.~3(g).
For $F_d>2.75$,
the labyrinth phase gradually 
collapses to another clump phase.
Figure~4(e) shows the corresponding $d\langle V_x\rangle/dF_{d}$ 
and $\sigma_{v}$ curves where the
transition to a more heterogeneous state at higher 
$F_d$ appears as an increase in $\sigma_{v}$. 
In the W state the initial depinning occurs via localized incommensurations
flowing along the pinning rows in one-dimensional channels while
the other vortices remain pinned, as shown
in 
Fig.~3(h) 
for $F_p=1.3$ and $F_{D} =0.96$. 
The vortices remain confined along the pinning rows
above the second depinning transition,
and as $F_d$ increases, the moving rows transition into the moving
labyrinth state shown in 
Fig.~3(i).
Fig.~4(f) shows two pronounced peaks in $d\langle V_x\rangle/dF_{d}$
for this sample, while
$\sigma_{v}$ increases at higher drives. 

For systems with purely repulsive vortex interactions 
there is generally at most only weak hysteresis in 
the velocity-force curves. 
For 
MI vortices,
we find hysteresis
for the strong pinning cases with both random and periodic pinning.
The double peak in
$d\langle V_x\rangle/dF_{d}$ is a very robust feature in these systems
for strong pinning and persists for larger system
sizes and various filling ratios $B_\phi/B$. 

In summary, we have studied vortex matter with long range attraction
and short range repulsion of the type proposed to occur in
type-I/type-II hybrid structures and multiband superconductors. 
In the absence of pinning, this system does not
form patterns or periodic arrays of stripes or bubbles, but instead
completely phase separates. 
In the presence of random or periodic substrates
we observe what we term a vortex wetting transition 
as a function of substrate strength or pinning density,
with a transition from a heterogeneous to a homogeneous vortex state 
for increasing substrate strength. 
Our results show that transport measurements can be used to test
for the existence of modified vortex interactions since,
under an applied drive, the strongly pinned
homogeneous states transition to highly heterogeneous states 
of moving stripes and clumps,
producing robust double peak features in the differential conductivity.  
Our results are also relevant to the broader class of systems with long 
range attraction
and short range repulsion on periodic and disordered substrates, which can 
be realized in 
a number of soft matter systems.

This work was carried out under the auspices of the 
NNSA of the 
U.S. DoE
at 
LANL
under Contract No.
DE-AC52-06NA25396.

\end{document}